\documentclass[fleqn,usenatbib]{mnras}

\usepackage{newtxtext,newtxmath}

\usepackage[T1]{fontenc}
\usepackage{ae,aecompl}

\usepackage{graphicx}    
\usepackage{amsmath}    
\usepackage{amssymb}    
\usepackage{hyperref}
\usepackage{caption}
\usepackage{subfigure}
\usepackage{ulem}
\usepackage{subfigure}

\newcommand{\kpc}{\,\mathrm{kpc}}
\newcommand{\magt}{\,\mathrm{mag}}

\newcommand{\Gyr}{\, \mathrm{Gyr}}
\newcommand{\FeH}{\, \mathrm{[Fe/H]}}

\title[Halo structure in SkyMapper]{Galactic cartography with SkyMapper: I.  Population sub-structure and the stellar number density of the inner halo}

\author[Z. Wan et al]{Zhen Wan$^{1}$\thanks{E-mail: zwan3791@uni.sydney.edu.au},
Prajwal R. Kafle$^{2}$,
Geraint F. Lewis$^{1}$,
Dougal Mackey$^{3}$,
\newauthor
Sanjib Sharma$^{1}$, and
Rodrigo A. Ibata$^{4}$
\\\\
$^{1}$Sydney Institute for Astronomy, School of Physics A28, The University of Sydney, NSW, 2006, Australia\\
$^{2}$ICRAR, The University of Western Australia, 35 Stirling Highway, Crawley, WA 6009, Australia\\
$^{3}$Research School of Astronomy \& Astrophysics, Australian National University, Canberra, ACT 2611, Australia\\
$^{4}$Observatoire de Strasbourg, 11, rue de l'Universit\' e, F-67000, Strasbourg, France\\
}

\pubyear{2017}

\begin{document}
\label{firstpage}
\pagerange{\pageref{firstpage}--\pageref{lastpage}}
\maketitle

\newcommand{\smp}{SkyMapper}

\begin{abstract}
The stars within our Galactic halo presents a snapshot of its ongoing growth and evolution, probing galaxy formation directly.
Here, we present our first analysis of the stellar halo  
from detailed maps of Blue Horizontal Branch (BHB) stars drawn 
from the \smp\ Southern Sky Survey.
To isolate candidate BHB stars from the overall population, we develop a machine-learning approach through the application of an
Artificial Neural Network (ANN), resulting in a relatively pure sample of target stars.
From this, we derive the absolute $u$ magnitude for the BHB sample to be $\sim2\ \magt$, varying slightly with $(v-g)_0$ and $(u-v)_0$ colours.
We examine the BHB number density distribution from 5272 candidate stars, deriving a double power-law with a break radius of $r_s = 11.8\pm0.3\ \kpc$, and inner and outer slopes of $\alpha_{in} = -2.5\pm0.1$ and $\alpha_{out} = -4.5\pm0.3$ respectively.
Through isochrone fitting of simulated BHB stars, we find a colour-age/metallicity correlation, with older/more metal-poor stars being bluer, and establish a parameter to indicate this age (or metallicity) variation. Using this, we construct the three-dimensional population distribution of BHB stars in the halo and identify significant substructure.
Finally, in agreement with previous studies, we also identify a systemic age/metallicity shift spanning $\sim3\kpc$ to $ \sim20\kpc$ in galactocentric distance.
\end{abstract}

\begin{keywords}
Survey: SkyMapper -- Star: Horizontal Branch -- Colour-Magnitude Diagrams -- Galaxies: Halo
\end{keywords}



\section{Introduction}
\label{sec:introduction}
Our own Milky Way presents us with the most detailed view of a large galaxy structure.
In his seminal work, \citet{Baade1954} initially proposed the hierarchical assembly of the Galaxy, picturing that our galaxy has accreted and merged with other galaxies to grow; several key pieces of evidence support this theory, such as the colour gradient in the globular cluster population \citep[e.g.][]{Searle1978}.
More recently, with high-resolution simulations have revealed the scars that hierarchical merging and accretion will leave on a galaxy, \citet{Bullock2005} suggests that our halo should be almost entirely built up from tidally disrupted systems. The different duration and time scale of accretion and merging---in the inner and outer parts of the halos---lead to a picture of younger, discrete substructures superimposed on an older, well-mixed background.
\citet{Font2006} suggests that the accretion history of galaxies stands behind the shape of the metallicity distribution of stars in the halo and in surviving satellites: the most-metal rich halo could have accreted a large number of massive satellites, indicating different grow up histories between the metal-rich M31 and the metal-poor Milky Way.

Observations could examine the conclusions from simulations. With the advance of imaging and spectroscopic technologies, a number of extensive surveys of the Galactic stellar halo have been undertaken such as: the Two Micron All Sky Survey \citep[2MASS;][]{Skrutskie2006}; Sloan Digital Sky Survey \citep[SDSS;][]{York2000}; the Sloan Extension of Galactic Understanding and Exploration \citep[SEGUE;][]{2009AJ....137.4377Y}; Large Sky Area Multi-Object Fibre Spectroscopic Telescope \citep[LAMOST;][]{Deng2012, Zhao2012} and the Canada-France Imaging Survey \citep[CFIS;][]{Ibata2017}.
Among targets in these observations, BHB stars have been proven to effectively indicate Galactic structure \citep[e.g.][]{2004AJ....127..899S,2006MNRAS.371L..11C,2010AJ....140.1850B,2011ApJ...738...79X,2011MNRAS.416.2903D}, kinematics \citep[e.g.][]{2004AJ....127..914S,2012ApJ...761...98K,2012MNRAS.422.2116K,2012MNRAS.425.2840D,2013MNRAS.430.2973K,2013ApJ...763L..17H} and dynamics \citep[e.g.][]{2008ApJ...684.1143X,2014ApJ...794...59K}.
These stars, having evolved off the main-sequence and possessing stable core helium burning, are bright, so that could be seen at large distance; are well-studied in absolute luminosity, so that could be pinned down in 3-dimension space \citep{2013MNRAS.430.1294F}. 
\citet{Preston1991}, an early relative work, found that the average colour of BHB stars shifts on order of $0.025\magt$ within galactocentric distances between $2\,\kpc$ to $12\,\kpc$, interpreting this shift as a gradient in age. This shift was examined and extend to $\sim 40\ \kpc$ later by \citet{Santucci2015}.
A recent research by \citet{Carollo2016} selected $\sim130,000$ BHB stars from SDSS to generate a high-resolution chronographic map reaching out to $\sim50~\kpc$. This large sample set enables them to identify structures such as the Sagittarius Stream, Cetus Polar Stream, Virgo Over-Density and others, and conclude that these are in good agreement with predictions from $\Lambda \mathrm{CDM}$ cosmology.

With the immense success of the northern hemisphere surveys, a complementary survey of the southern sky become essential.
To create a deep, multi-epoch, multi-colour digital survey of the entire southern sky, the \smp\ project was initiated \citep{2018arXiv180107834W}. Recently, the first all-sky data release (DR1)---covering $20,200\,\rm{deg}^2$ of the sky, with almost 300 million detected sources, including both stellar and non-stellar sources---has been published.
In this paper, we make use of the \smp\ Southern Sky Survey to provide the first detailed population and number distribution maps of BHB stars, exploring and examining the properties of the Galactic halo.

The paper is structured as follows; Sec.\ref{sec:data} presents details of the data selection, outlining our approach to the identification of BHB stars from colour-colour relationships drawn from the \smp\ DR1 Southern Sky Survey.
In Sec.\ref{sec:Distance_Cal}, we present the absolute magnitude calibration of our BHB sample based upon several stellar clusters and parallax separately.
We discuss the correlation of BHB stellar ages and metallicity with their \smp\ colour in  Sec.\ref{sec:age_cal}, and finally, present our southern sky colour/population map and BHB number density distribution in Sec.\ref{sec:result}.
In closing, Sec.\ref{sec:conclusion} summarises our results and conclusions.

\section{Data}\label{sec:data}
\subsection{Observational Data}
The \smp\ filter set has 6 bands, with its $r, i$ and $z$ bands being similar to corresponding SDSS bands. Its $u$ band is relatively bluer than the corresponding SDSS filter, emulating the $Str \ddot{o} mgren\ u$ band, while the $g$ band is redder than the SDSS $g$ band. Between these redefined $u$ and $g$ bands, \smp\ has a relatively narrow $v$ band, which is metallicity-sensitive \citep{Bessell2011}. In its DR1, \smp\ has reached limits of $17.75, 17.5, 18, 18, 17.75, 17.5 \magt$ for its $u, v, g, r, i$ and $z$ band photometry respectively. 

The aim is to find the BHB stars in \smp\ DR1 as much as possible, with the contamination kept as less as possible. Blue Straggler (BS) stars are the primary contamination in BHB sample set selected by colour \citep[e.g.][and references therein]{2002MNRAS.337...87C,2004AJ....127..899S}, which cannot be ignored, considering that the ratio of BS stars to BHB stars is $\sim 1.5-2.0$ in the inner halo \citep{Santucci2015b}. Additionally, though less significantly, the hot Main Sequence (MS) stars could be mixed up in the selected BHB sample (see later in the colour-colour diagram).
To distinguish BHB and BS, as well as MS stars, some earlier works, such as \cite{2002MNRAS.337...87C} and \cite{2004AJ....127..899S}, presented a robust method based upon the Balmer $H_{\gamma}$ and $H_{\delta}$ line profile parameters to cull the contaminants. Though this is not feasible to \smp\ since it only has the photometric measurement, we applied the method to SEGUE spectrum data and acquired a clean \smp\ BHB/BS/MS sample by crossmatch in the overlap region of SEGUE and \smp. This sample is then used as a training set to help train an ANN to locate the BHB stars in the \smp\ colour-colour diagram, which were exerted on the entire \smp\ DR1 catalogue to isolate candidate BHB stars.

We note that the extinction has been corrected throughout based upon the \citet[][hereafter S98]{Schlegel1998}\footnote{S98 are known to overestimate the amount of reddening, and we exclude the low galactic latitude stars where the extinction is significant. For our sample, the difference of E(B-V) of our sample predicted by the more recent \citet[][ hereafter S10]{Schlafly2010} and S98 is less than $0.06 \magt$ (mostly less than $0.02 \magt$) and we employ S98 for consistency with previous studies.}.

\subsection{SEGUE BHB Stars}
\label{sec:BHB_From_SEGUE}
\begin{figure}
\centering
\begin{subfigure}
	\centering
    \includegraphics[width=\columnwidth]{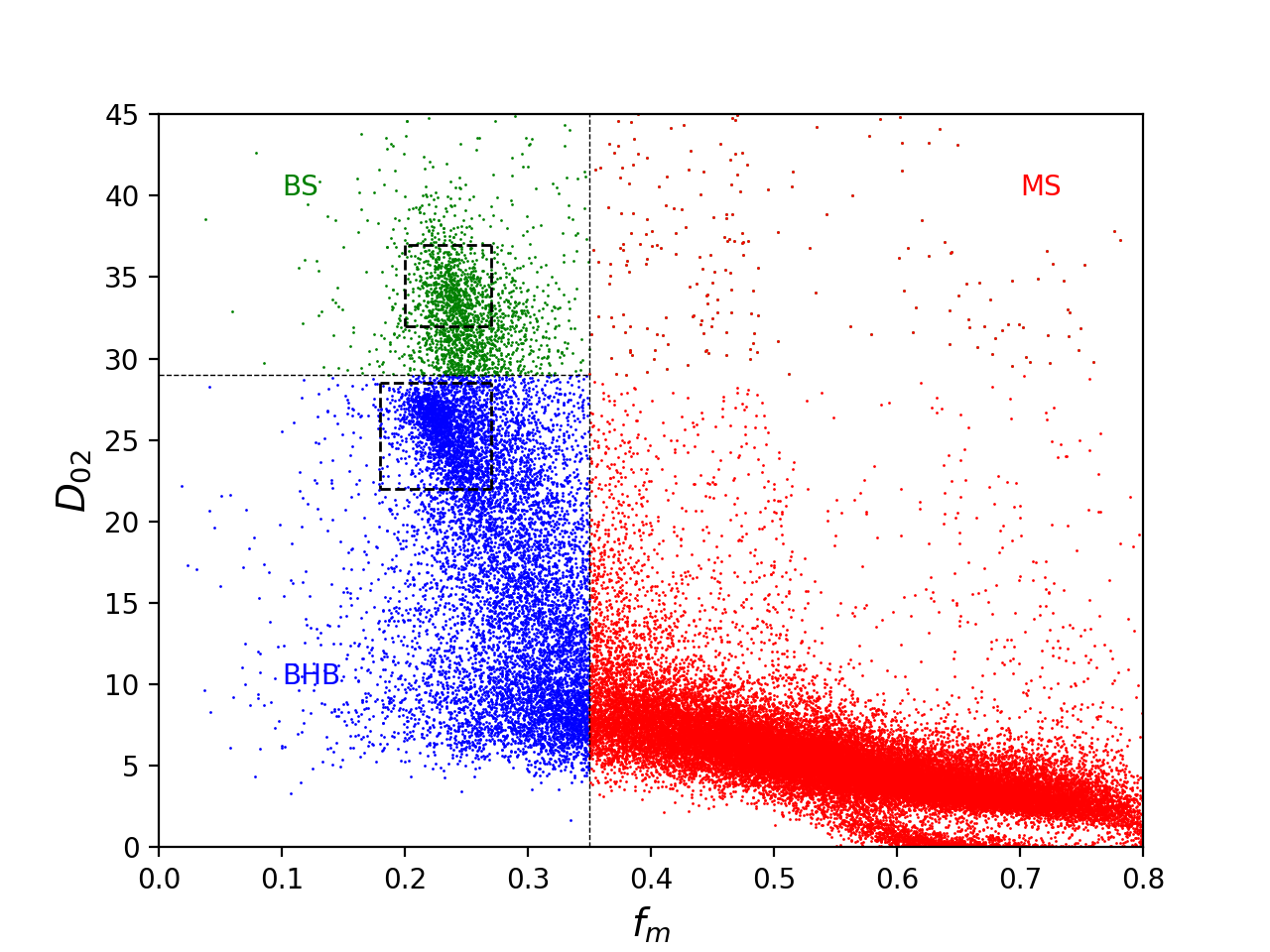}
\end{subfigure}
\begin{subfigure}
	\centering
    \includegraphics[width=\columnwidth]{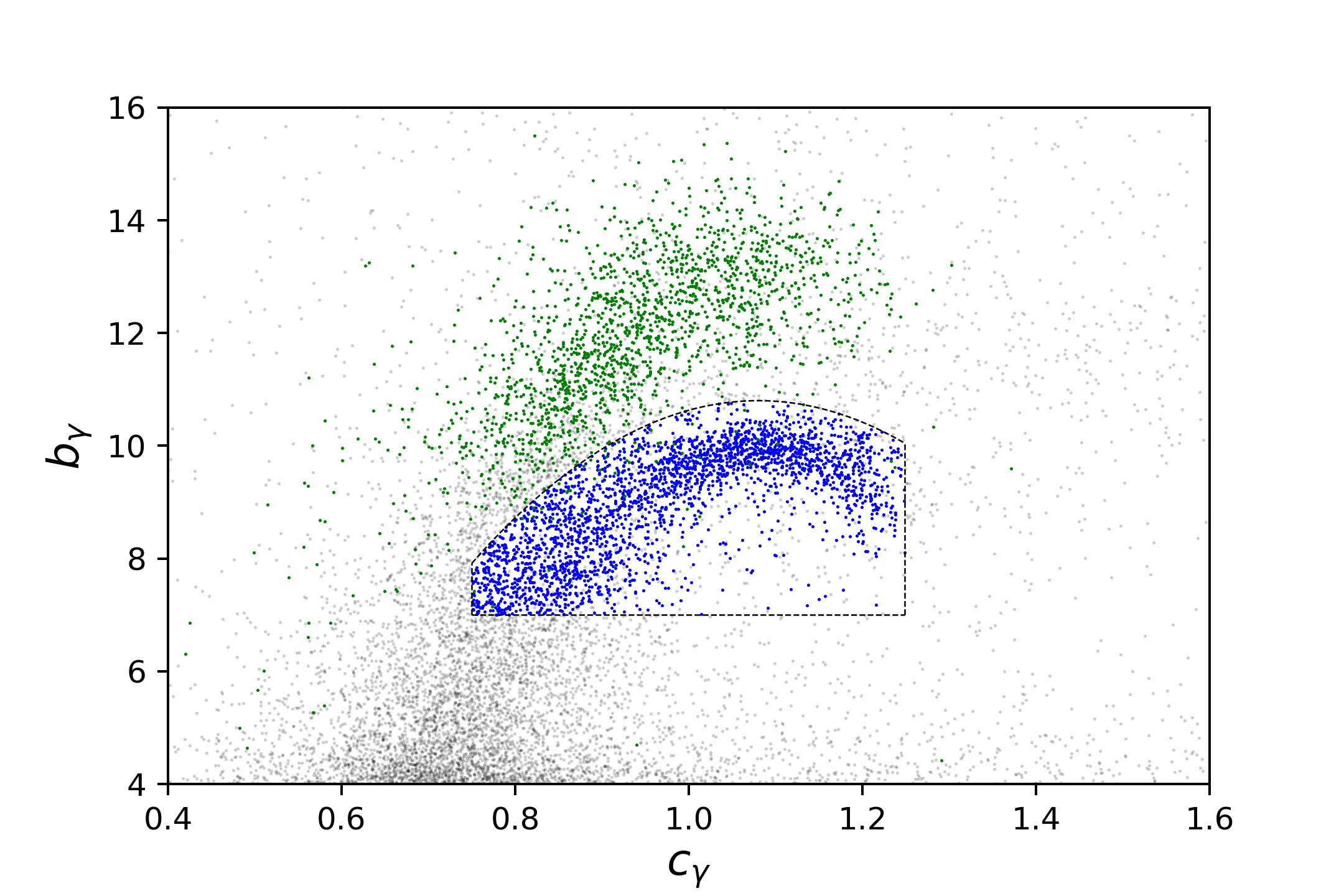}
\end{subfigure}
\caption{The line profile parameters extracted from SEGUE with the BHB selection criteria (thin dash lines) from \citet{2004AJ....127..899S}. In each figure, we colour the BHB stars that meet the criteria blue; colour the BS stars green and the MS star red. {\it Upper}: The distribution of $D_{0.2}$ and $f_m$ of H$_{\delta}$ line (see the definition in Sec.\ref{sec:BHB_From_SEGUE}). The thick dash lines divide the figure into three parts: BHB, BS and MS regions. We use two thin-dash-line boxes to indicate where BHB and BS stars concentrate. {\it Bottom}: The distribution of H$_{\gamma}$ line profile parameters: $c_{\gamma}$ and $b_{\gamma}$. The dash lines outline the region where BHB stars locate, which is clearly isolated from BS stars. The grey points represent MS stars.}
\label{fig:linecut}
\end{figure}

SEGUE (including SEGUE-1 and SEGUE-2) library contains more than 300,000 spectra, which makes it impossible to profile all stars with limit computational time; hence, before we could select BHB sample from SEGUE, we apply a geometry and a colour cuts to reduce the sample size. In detail, a geometry cut:
\begin{equation}
Dec. < 15^{\circ}
\end{equation}
is applied first, representing the overlap region between \smp\ and SEGUE. Then we select star within:
\begin{gather}
    0.80<(u-g)_0<1.35 \notag \\
    -0.40<(g-r)_0<0.00
\end{gather}
to roughly select BHB stars. These two cuts reduce the SEGUE spectra size to $\sim 50,000$ stars, on which we exert the line profile selection from \citet{2004AJ....127..899S}. We use S\'{e}rsic profile to depict H$_{\gamma}$ and $H_{\delta}$ lines by:
\begin{equation}
    y = 1 - A\ exp\left[-\left(\frac{|\lambda - \lambda _0|}{b}\right)^c\right]
    \label{eq:sersic}
\end{equation}
where $A, b$ and $c$ are three key parameters describe the line shape; $y$ is the normalized line intensity; $\lambda$ and $\lambda_0$ represent wavelength and central wavelength. Additionally, two more quantities are defined: $D_{0.2}$ is the width of Balmer line at $20\%$ below the local continuum, and $f_m = 1 - A$ is the relative flux at the line core. We extract the H$_{\gamma}$ and $H_{\delta}$ lines profile of all the remained SEGUE stars with Markov Chain Monte Carlo \citep[MCMC;][]{2013PASP..125..306F}, and apply the cuts:
\begin{gather}
    17 \le D_{0.2}/\text{\AA} \le 28.5  \notag \\
    0.1 \le f_m \le 0.3
\end{gather}
on $H_{\delta}$ lines profile and:
\begin{gather}
    0.75\le c_{\gamma} \le 1.25 \notag \\
    7.50 \mathring A \le b_{\gamma} \le 10.8-26.5\left(c_{\gamma} - 1.08\right)^{2}
\end{gather}
on H$_{\gamma}$ line profile to select an clean BHB sample set. \autoref{fig:linecut} demonstrates the line profile parameters distribution and the selection we applied, with which we finally end up with 2610 SEGUE BHB stars, as well as 1999 BS stars, and 43281 MS stars.

\begin{table*}
    \centering
    \caption{An example table of the BHB/BS/MS sample set drawn from \smp. The first column is the \smp\ object ID, which can be used to locate stars. The second column is the corresponding Spectral ID in SEGUE, in the format of PLATE-MJD-FIBER. The third and fourth columns give the location of the star in J2000 from \smp. The fifth to tenth column presents the \smp\ $u, v$ and $g$ bands photometric result and uncertainty, which are used as an indicator for classification. The last column is the star type as determined as part of this study.
    }
    \label{tab:training set}
    \begin{tabular}{*{11}{c}} 
        \hline
        Object id & SDSS SPECID & R.A. & Dec. & $u_{psf}$ & $e_{u_{psf}}$ & $v_{psf}$ & $e_{v_{psf}}$ & $g_{psf}$ & $e_{g_{psf}}$ & Type\\
        \hline
        68139182 & 0513-51989-0528 & 174.654 & 3.149 & 16.602 & 0.009 & 15.820 & 0.005 & 15.133 & 0.014 & BHB \\
        47695470 & 2057-53816-0126 & 124.320 & -0.469 & 15.475 & 0.003 & 14.835 & 0.012 & 14.222 & 0.010 & BHB \\
        67098280 & 2716-54628-0474 & 209.700 & -8.977 & 14.988 & 0.024 & 14.064 & 0.007 & 13.683 & 0.006 & BHB \\
        47066082 & 2806-54425-0124 & 122.583 & -8.684 & 15.014 & 0.006 & 14.395 & 0.027 & 13.819 & 0.021 & BHB \\
        47066563 & 2806-54425-0259 & 122.251 & -8.599 & 15.596 & 0.026 & 14.861 & 0.019 & 14.199 & 0.018 & BHB \\
        46953790 & 2806-54425-0378 & 121.941 & -6.800 & 16.541 & 0.020 & 15.744 & 0.007 & 15.142 & 0.009 & BHB \\
        100270164 & 0922-52426-0571 & 231.901 & -1.435 & 17.774 & 0.035 & 16.961 & 0.078 & 16.439 & 0.021 & BS \\
        100910930 & 0924-52409-0281 & 236.479 & -1.223 & 17.158 & 0.042 & 16.527 & 0.055 & 15.828 & 0.005 & BS \\
        5347974 & 0926-52413-0408 & 335.443 & -0.123 & 18.089 & 0.243 & 17.712 & 0.026 & 17.092 & 0.013 & BS \\
        48442794 & 0984-52442-0259 & 129.310 & 4.299 & 16.471 & 0.029 & 15.870 & 0.028 & 15.156 & 0.006 & BS \\
        48434506 & 0990-52465-0452 & 129.713 & 3.896 & 16.644 & 0.014 & 15.838 & 0.012 & 15.294 & 0.002 & BS \\
        65406019 & 2178-54629-0442 & 182.693 & -0.233 & 16.254 & 0.011 & 15.916 & 0.011 & 15.306 & 0.009 & MS \\
        65396598 & 2178-54629-0512 & 181.608 & -0.566 & 18.110 & 0.054 & 17.478 & 0.052 & 16.848 & 0.006 & MS \\
        65393432 & 2186-54327-0437 & 180.421 & -0.652 & 16.958 & 0.027 & 16.628 & 0.034 & 15.715 & 0.004 & MS \\
        68375791 & 2198-53918-0007 & 182.443 & 0.938 & 16.047 & 0.017 & 15.729 & 0.011 & 14.945 & 0.004 & MS \\
        65430024 & 2198-53918-0027 & 184.064 & -0.295 & 17.640 & 0.018 & 17.378 & 0.078 & 16.427 & 0.008 & MS \\
        \vdots & \vdots & \vdots & \vdots & \vdots & \vdots & \vdots & \vdots & \vdots & \vdots & \vdots \\
        \hline
    \end{tabular}
\end{table*}

\begin{figure*}
	\includegraphics[width=\textwidth]{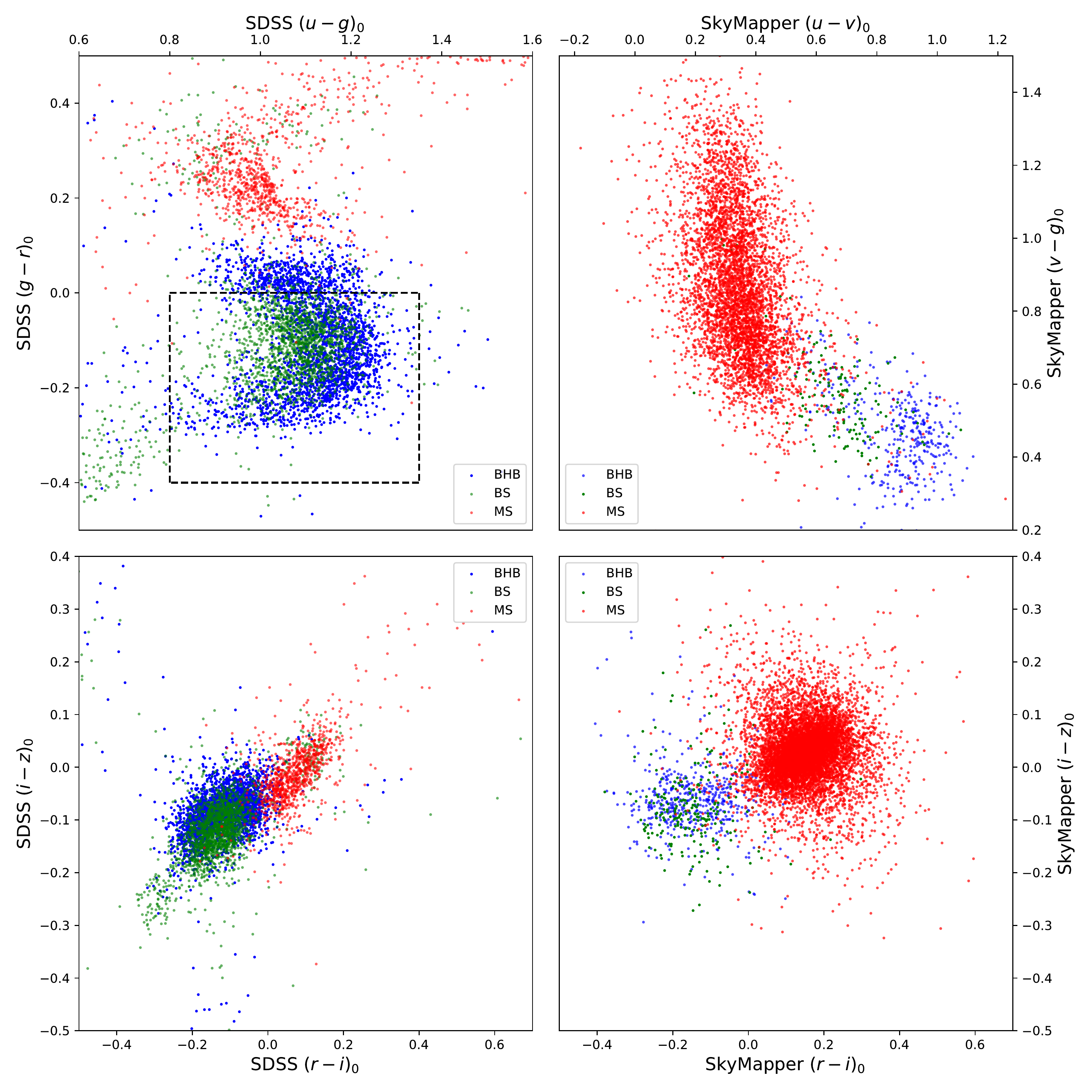}
        \caption{Comparison of the BHB/BS/MS colour-colour distributions for SDSS and \smp\ photometry
focussing upon $(g-r)_0$ versus $(u-g)_0$ and $(i-z)_0$ versus $(r-i)_0$. 
The dashed rectangle in the upper-left panel indicates the colour cut that selects BHB stars in the SDSS colour-colour diagram. In the upper panels, we compare two blue colours. BHB stars and BS stars are more clearly separated in \smp\ colours. The bottom panels present the distribution of BHB/BS/MS stars in two red colours. 
Clearly, in the red colours, the BHB and BS stars are indistinguishable.}
        \label{fig:sample_sm_colour}
\end{figure*}

\subsection{Training the ANN to identify BHB stars from \smp}
\label{sec:training_ANN}
Before we train the ANN and find all the BHB stars in \smp\ DR1, we crossmatch between \smp\ and SEGUE with the position difference less than $1.8\arcsec$, from which we find 374 BHB stars, 156 BS stars, and 4917 MS stars in \smp. Here in \autoref{fig:sample_sm_colour}, we compare the colour-colour diagrams of BHB/BS/MS stars from \smp\ and SDSS. \smp\ colours, $(u-v)_0$ vs $(v-g)_0$ and $(r-i)_0$ vs $(i-z)_0$, are presented; SDSS colours, $(u-g)_0$ vs $(g-r)_0$ and $(r-i)_0$ vs $(i-z)_0$, are shown as comparison. For both \smp\ and SDSS, we find that BHB stars are indistinguishable from BS stars within red colours diagrams; in the blue colours diagrams, the BHB stars are more clearly isolated from BS stars in \smp\ than in SDSS---due to the difference between \smp\ and SDSS filter sets---with which we expect to train a reliable ANN.

We randomly select $5\%$ of our sample as a test set, which contains 19 BHB stars, 8 BS stars and 246 MS stars; the training set, with test set excluded, has 355 BHB stars, 148 BS stars, and 4671 MS stars. 
Then we apply a 2-hidden-layer ANN with 32 neurons---an algorithm from scikit-learn \citep{scikit-learn} with multi-layer perceptron---with $(u-v)_0$ and $(v-g)_0$ as indicators, which, as in \autoref{fig:sample_sm_colour}, depict the boundary of BHB stars clearly.

\begin{figure}
    \subfigure{\includegraphics[width=\columnwidth]{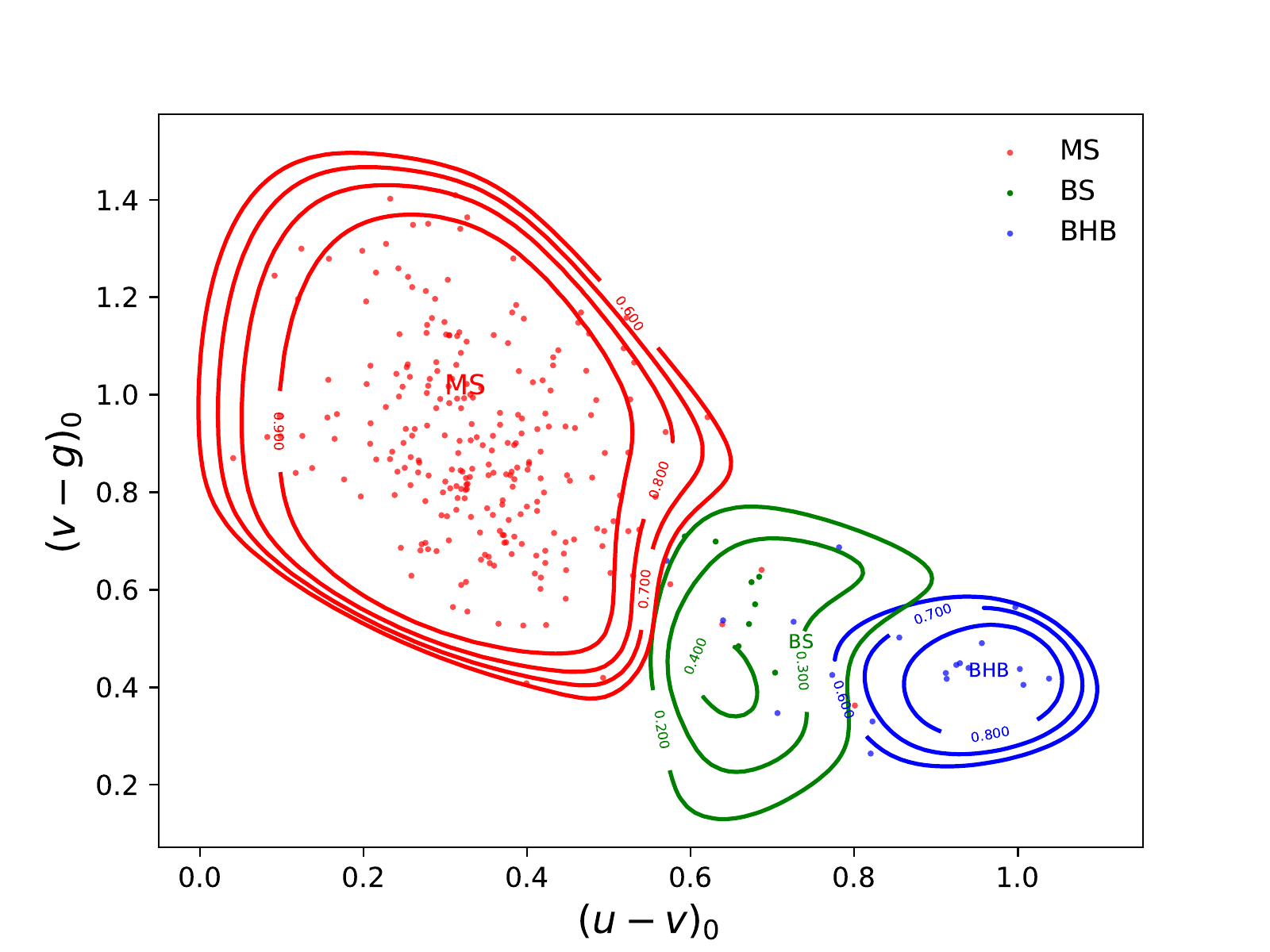}}
    \subfigure{\includegraphics[width=\columnwidth]{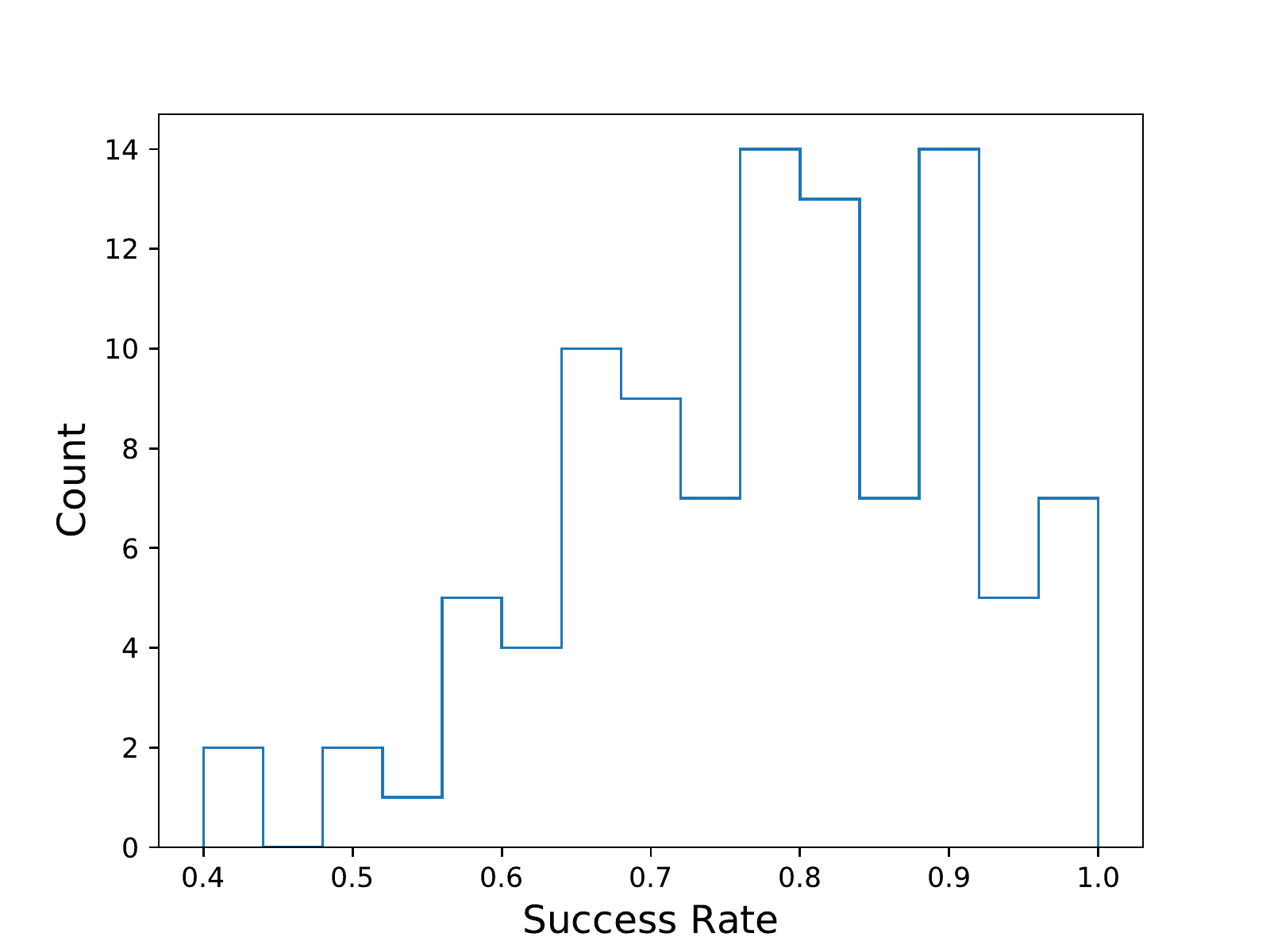}}
        \caption{{\it Upper}: The probabilities for each type overplotted on the scattered test set, where BHB stars are in blue. BS stars are in green, and MS stars are in red. We adopt the resulting classification if the probability is larger than $60\%$. Several BHB stars in the test set reach into the to BS/MS region, but the adopted BHB region is relatively clean. {\it bottom}: 
The success rate of the successful identification of stars as the result of 100 tests with the ANN and different test sets and training sets. This success rate peaks at 0.8.
}
        \label{fig:ANN_test}
\end{figure}

When fitting the training set, we assume the probability of a star being BHB star to be zero when the star is located far enough from our sample in colour-colour diagram. The top panel of \autoref{fig:ANN_test} shows the ANN fitting result, where the probabilities being BHB/BS/MS stars are drawn as contours. We concerned that the selection effect may influence the fitting considering the size of training/test set; to avoid that, we perform the fitting 100 times with randomly selected training/test set. The bottom panel of \autoref{fig:ANN_test} demonstrates the ratio of successfully identified BHB in each test, which tells us that most tests have the success rate higher than $80\%$. Later in this study, we select a star if its probability of being a BHB star is larger than $60\%$, providing good agreement with predicted distributions. 

\begin{figure}
    \includegraphics[width=\columnwidth]{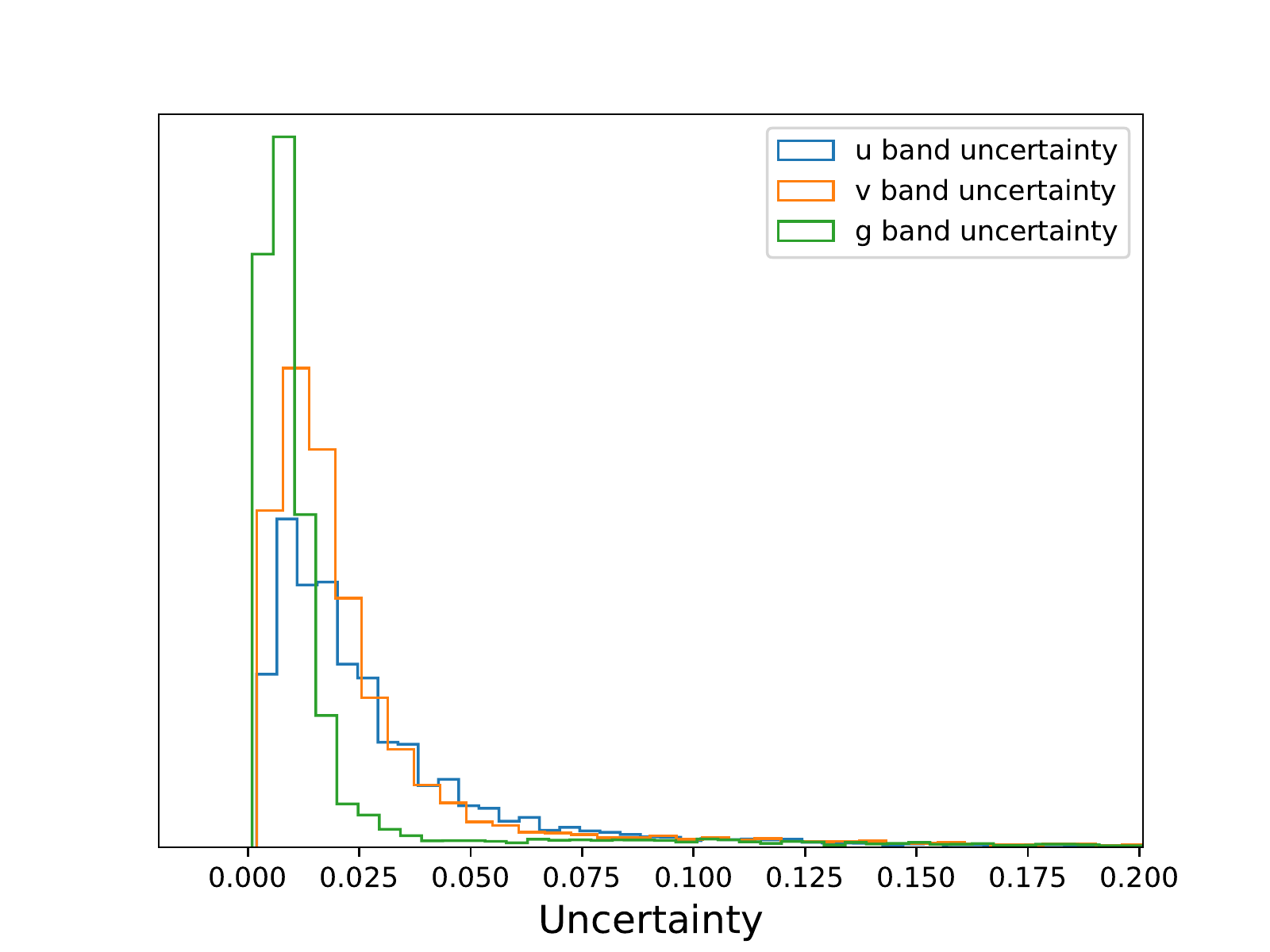}
        \caption{Normalised uncertainty distribution for photometry in the $u, v, g$ bands. Most stars in our data sample have photometric uncertainty smaller than $0.1 \magt$. }
        \label{fig:ErrorDistribution}
\end{figure}

When preparing the \smp\ data, we select stars whose latitude $\left|b \right|$ is larger than $25^\circ$ to avoid contamination from the Galactic disk. Additionally, following photometric quality labels are considered:
\begin{gather}
nimaflags = 0, \notag \\  
flags = 0, \notag \\
ngood > 1, \notag \\
ngood\_min > 1\ and\notag \\
nch\_max = 1
\end{gather}
to ensure a clean and reliable dataset. The photometric uncertainties in the $u, v$ and $g$ bands are shown in \autoref{fig:ErrorDistribution}, which clearly shows that most stars in our sample have photometric uncertainties smaller than $0.1 \magt$.

\begin{figure}
    \includegraphics[width=\columnwidth]{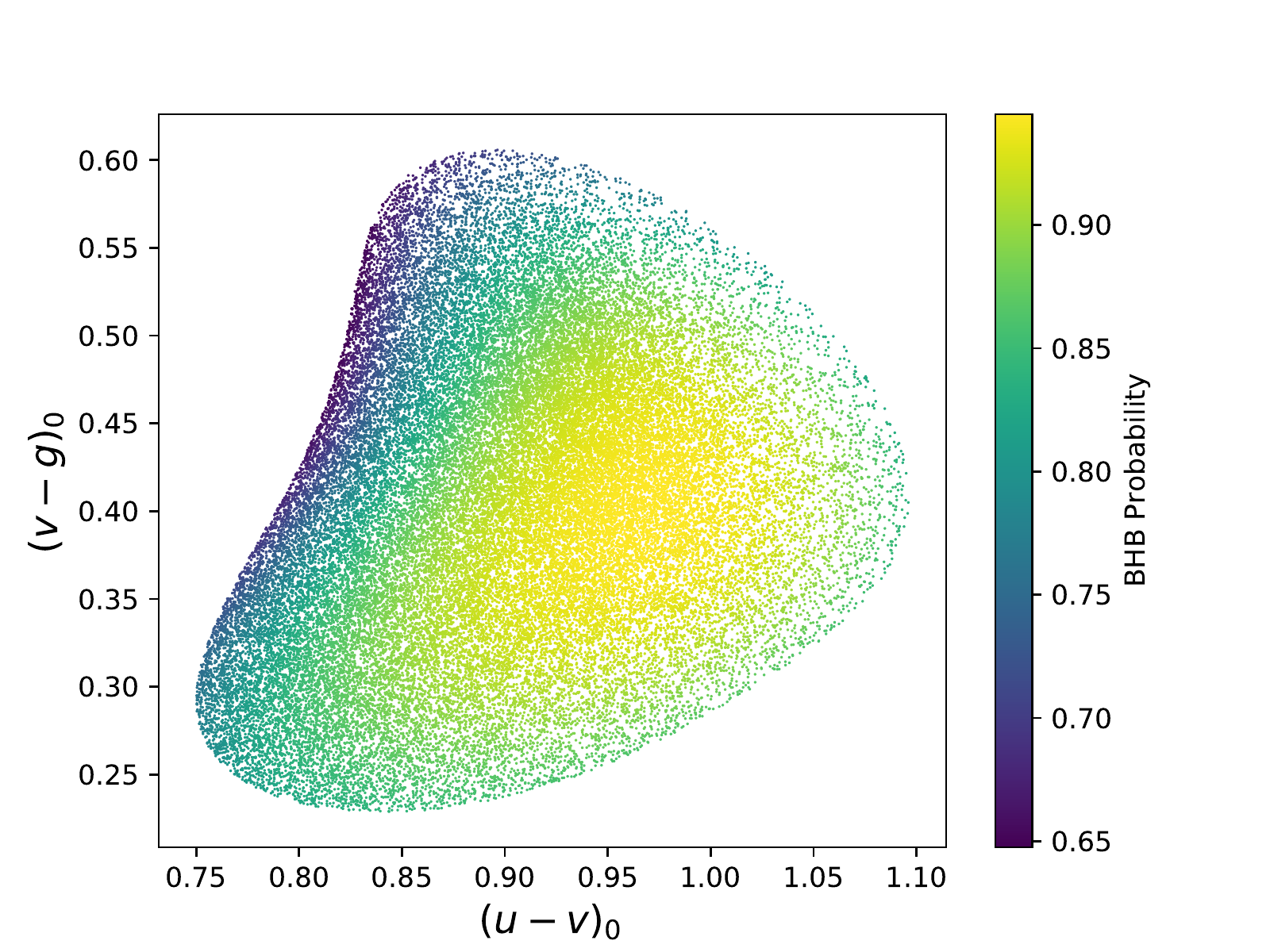}
        \caption{BHB stars from \smp\  selected by our trained ANN. The probability of each stars being BHB star is presented by colour. Stars are concentrated at $((u - v)_0, (v - g)_0)$, where the probability of being BHB stars is the highest.}
        \label{fig:ann_selected}
\end{figure}

In total, the ANN selected 16970 BHB stars from \smp\ DR1. We present their colour distribution in \autoref{fig:ann_selected}, where probabilities of the stars being BHB stars are colour coded. The training set and selected BHB stars are available online as supplementary material.

\section{Distance Calibration}\label{sec:Distance_Cal}
BHB stars are wildly used as standard candle (see references in Sec.\ref{sec:introduction}); the actual absolute magnitude of each BHB star is influenced by the properties of the stellar envelope, particularly the metallicity and temperature \citep[e.g.][]{Wilhelm1999, 2004AJ....127..899S, 2013MNRAS.430.1294F}, which also vary the colour of BHB stars. 
We measure BHB stars absolute magnitude in \smp\ photometry with two independent methods:
calibrate the absolute magnitude with stellar clusters; and with {\it Gaia} DR1 parallax \citep{Gaia2016a,  Gaia2016b}. Following in this section, we present the details of each method.

\subsection{Scaling relation from Clusters}
The distances to globular clusters are reliable due to many aspects: the globular clusters are compact so that the distance dispersion per system is negligible, and varied distance measurements are feasible to globular clusters like by RR Lyrae stars and isochrone fitting. The trustful distance inspires us to calibrate the absolute magnitude of BHB stars with well studied globular clusters.

For this study, we choose four fiducial globular clusters as distance calibrators: NGC5139 ($\omega$ Centauri), M22, NGC6397 and M55.
They are all located at roughly $\sim 10\ \kpc$ with respect to the Sun, each possessing well-measured distance moduli.
Here we adopt four latest work on the distance moduli of the above four clusters: \citet{Braga2018} use RR Lyrae stars to constrain NGC5139's distance and extinction; \citet{Kunder2013} also use RR Lyrae stars for M22; \citet{McNamara2011} takes the mean of the results from RR Lyrae stars and $\delta$ Scuti for M55; \citet{Brown2018} calibrates the distance for NGC 6397 through trigonometric parallax.
Table \ref{tab:globlar_cluster} presents the resultant extinction and distance moduli of the four clusters used in this study.

\begin{table}
    \caption{The four globular clusters used as the distance calibrators in this study. The first column presents their name, while the second
    is their extinction  and the last column is their distance modulus. The distance to NGC 5139 ($\omega$ Centauri), M22 and M55 are
    calculated based on RR Lyrae stars where NGC 6397 is based on main sequence fitting. The distance moduli (DM) are extinction corrected.}
    \label{tab:globlar_cluster}
    \begin{tabular}{*{4}{c}}
        \hline
        Name & $E(B-V)_{\it SFD}$ / $\magt$ & DM / $\magt$\\
        \hline
        NGC5139 & $0.13\ $ & $13.67 \pm 0.04\ $ \\
        M22 & $0.32\ $ & $12.67 \pm 0.12\ $ \\
        NGC6397 & $0.19\ $ & $11.89 \pm 0.11\ $ \\
        M55 & $0.14\ $ & $13.62 \pm 0.05 $ \\
        \hline
    \end{tabular}
\end{table}

\begin{figure}
    \includegraphics[width=\columnwidth]{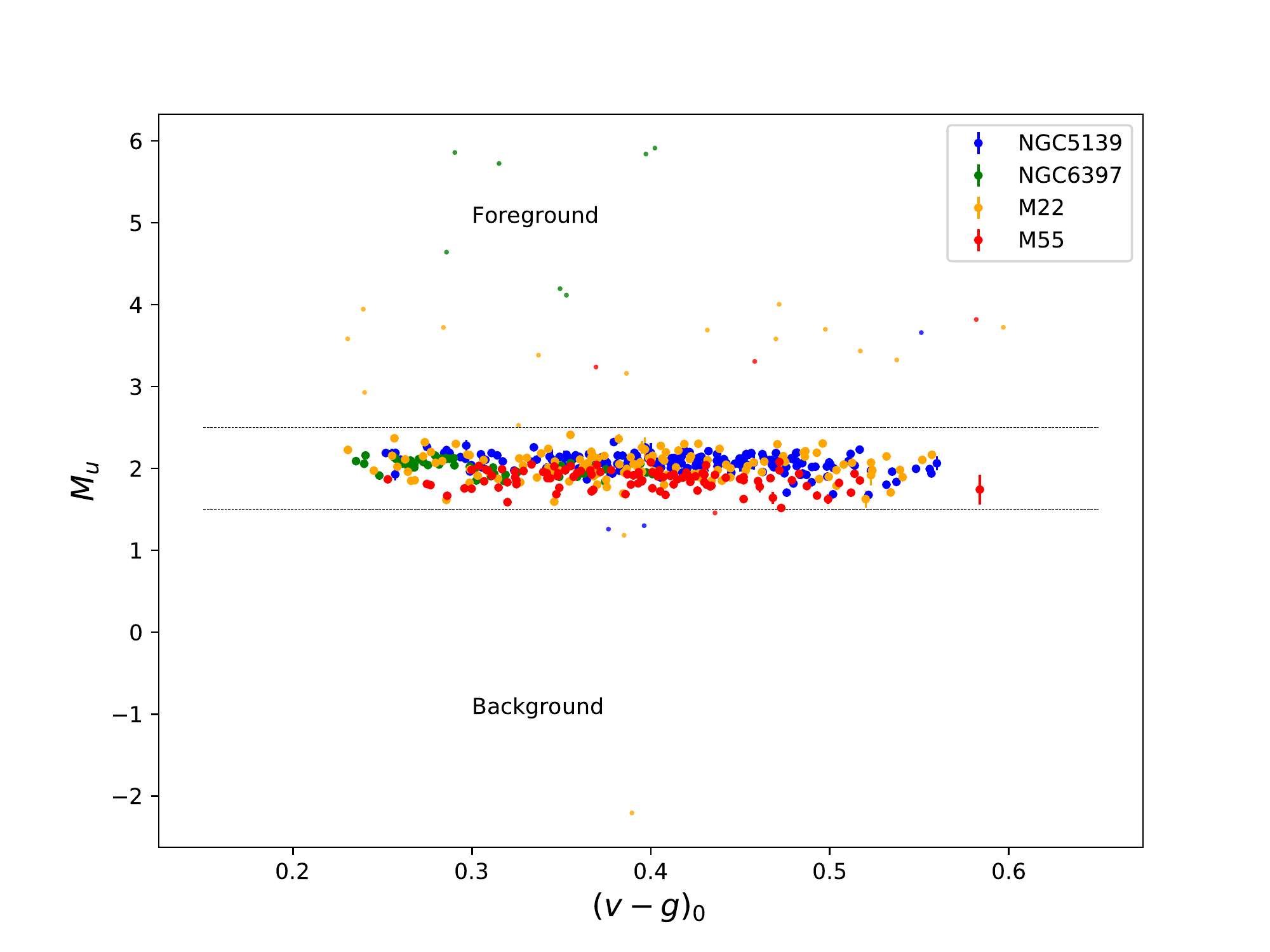}
        \caption{The CMD of stars in NGC5139 (blue), NGC6397 (green), M22 (orange) and M55 (red) with uncertainties marked.  Stars that belong to these clusters concentrate around $M_u = 2 \magt$, and foreground and background stars sit distinct from this sequence.}
        \label{fig:FourClusters}
\end{figure}

We employ the trained ANN to select BHB stars within $1^\circ \times 1^\circ$ regions that centred at each cluster. The crowding issues could be crucial at the centre of globular clusters, which we avoid by requiring the PSF photometric quality being better than 0.9 so that most stars in the centre of the clusters are ignored.
Following this, we applied the adopted distance moduli to each field to calculate the absolute magnitude of our sample and assemble them in \autoref{fig:FourClusters}, where the BHB stars concentrate around $M_u = 2 \magt$, with some foreground and background stars dispersed.

\begin{figure}
    \includegraphics[width=\columnwidth]{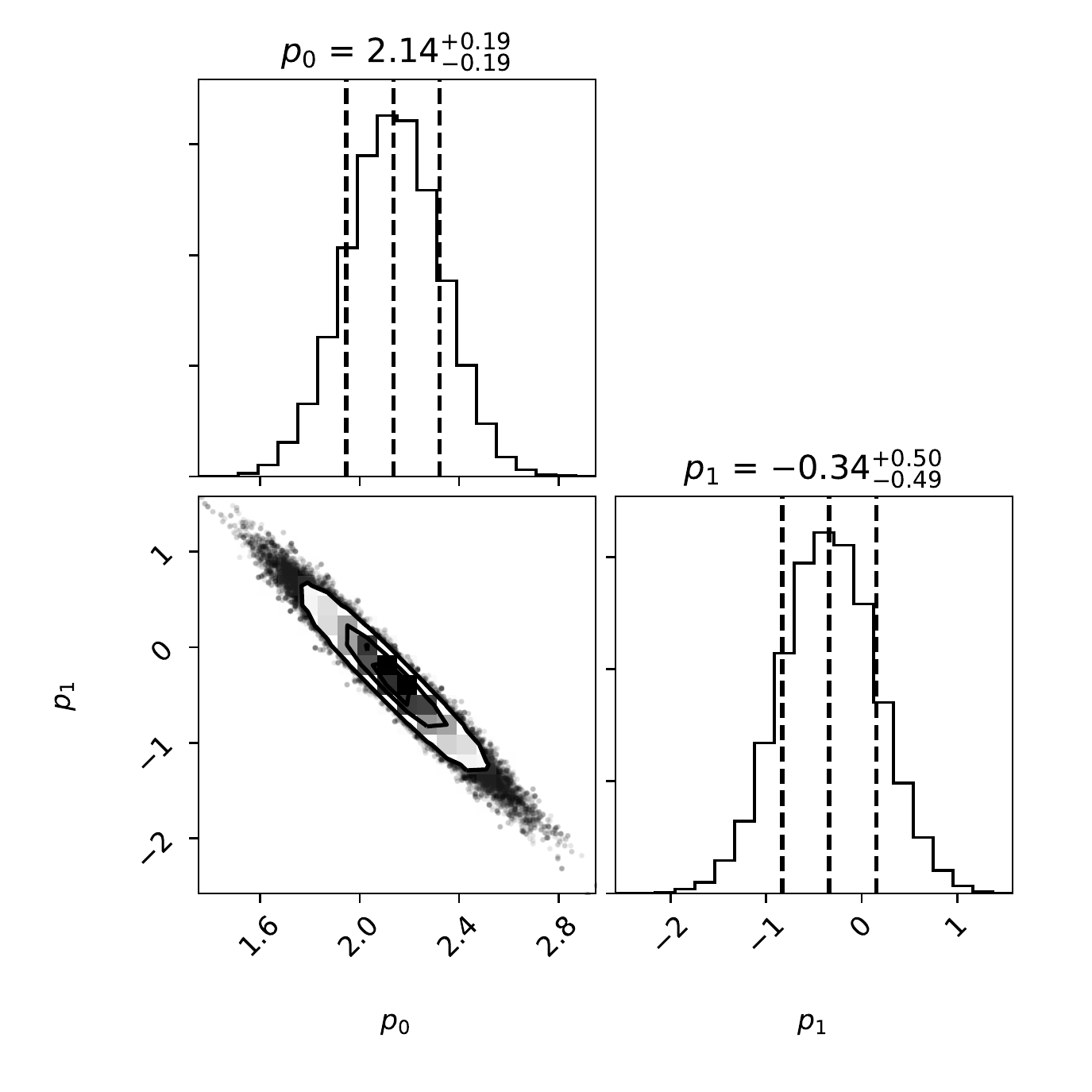}
        \caption{The colour - absolute magnitude fitting result. The most likely values and their $1\ \sigma$ ranges are marked in titles.}
        \label{Fig:AbsoluteMagnitudeFitting}
\end{figure}

The absolute magnitude in \autoref{fig:FourClusters} varies slowly with colour, which is what we expect for old BHB stars. To describe that, we use a MCMC \citep{2013PASP..125..306F} routine (see the parameter probability distribution in \autoref{Fig:AbsoluteMagnitudeFitting}) to fit the distribution with a simple linear function:
\begin{gather}
    M_u = p_0 + p_1\times (v-g)_0 \notag \\
    p_0 = 2.14^{+0.19}_{-0.19},\ p_1 = -0.34^{+0.50}_{-0.49}
    \label{eq:linear_part}
\end{gather}
Here in this fitting, we exclude foreground and background stars by selecting stars between $1.5 \le M_u \le 2.5$, which is a broad cut to encompass the entire BHB population as shown in \autoref{fig:FourClusters}.

\begin{figure}
 	\includegraphics[width=\columnwidth]{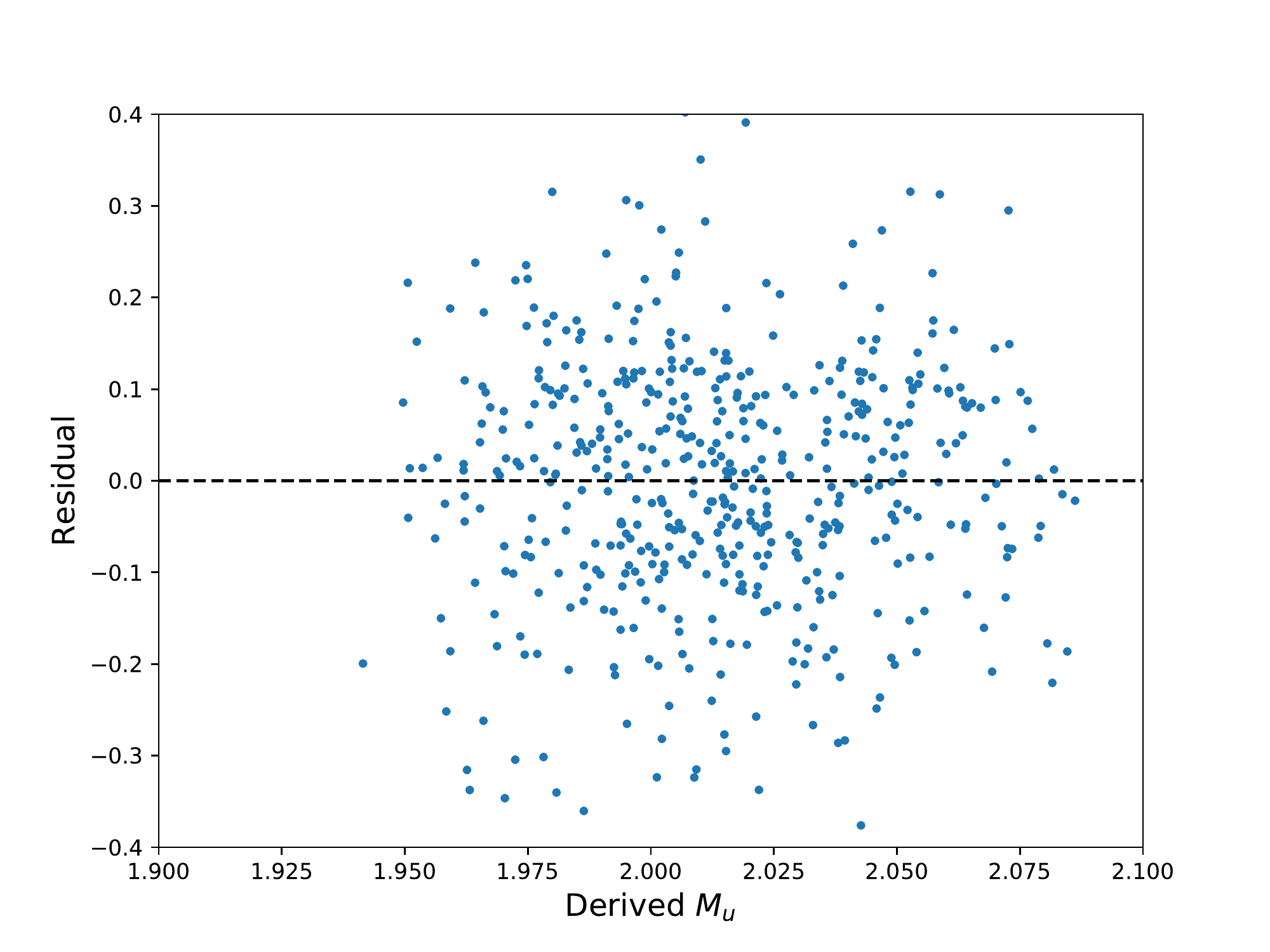}
	\caption{The difference of inferred absolute magnitude from clusters and the fitted absolute magnitude. The majority of the deviations are less than $0.2 \magt$. }
	\label{fig:deviation}
\end{figure}

We compare the absolute magnitude inferred from this fitting result with the original value from cluster calibration, and in \autoref{fig:deviation}, we present the residual against inferred absolute magnitudes. The {\it rms} of the residual is $0.124 \magt$, corresponding to a distance uncertainty of $5\%$. We assume the intrinsic uncertainties---like the actual properties and environment of each star, the size of the globular clusters---are included in above $5\%$ distance uncertainty; additionally, we mulled over the clusters distances uncertainty due to differing approaches and found it beyond the scope of this paper. In considering the potential influence on our the final results, we adopt another $5\%$ systematic distance error (corresponding to $0.124 \magt$, assuming they are at the same scale of the {\it rms}) into our final calculations.

\begin{figure}
  \includegraphics[width=\columnwidth]{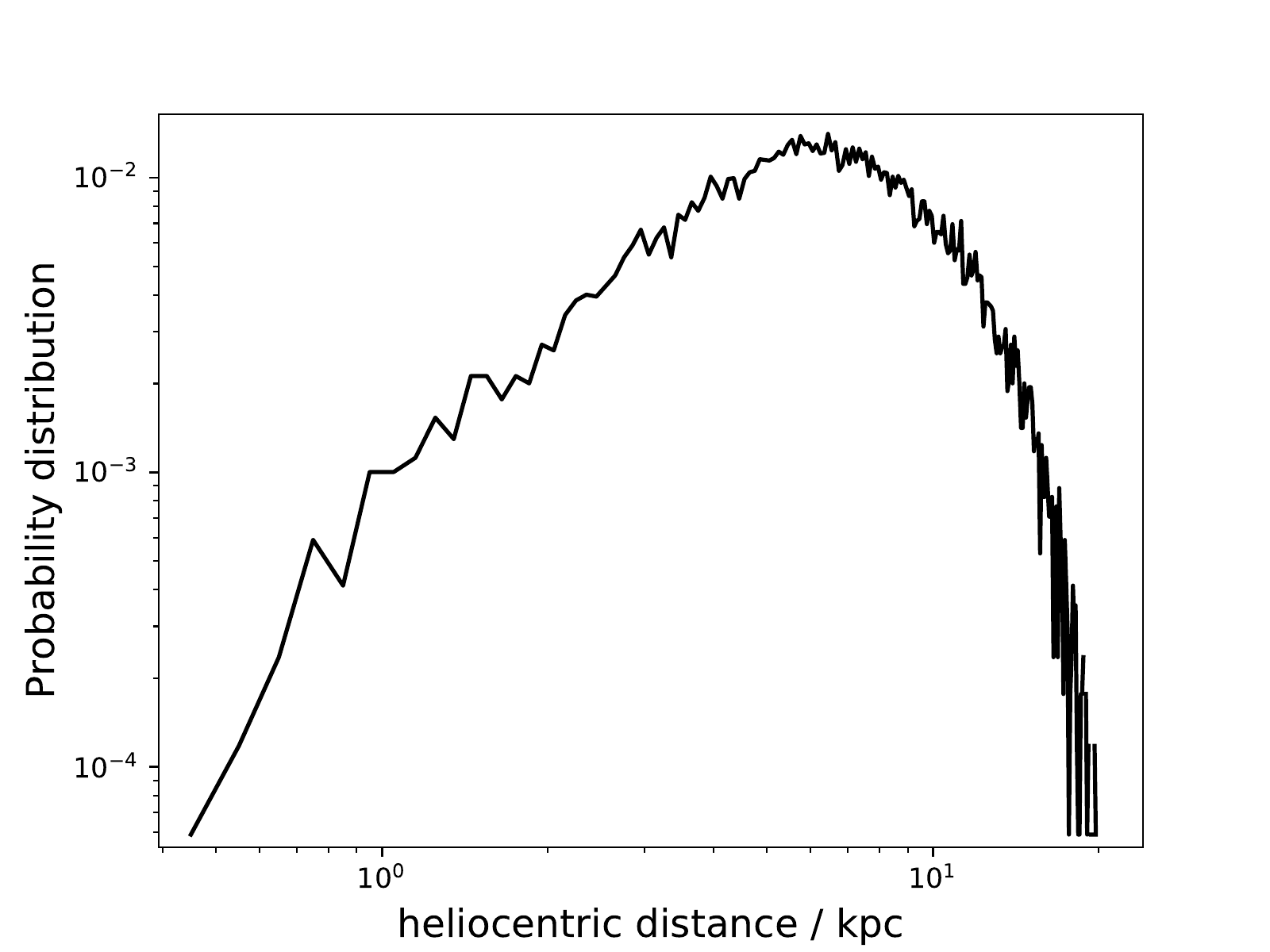}
  \caption{The number distribution of BHB stars from our sample in logarithm space. The sample covers heliocentric distance $r < 11\ \kpc$. }
  \label{Fig:Numberdistribution}
\end{figure}

\begin{figure}
  \includegraphics[width=\columnwidth]{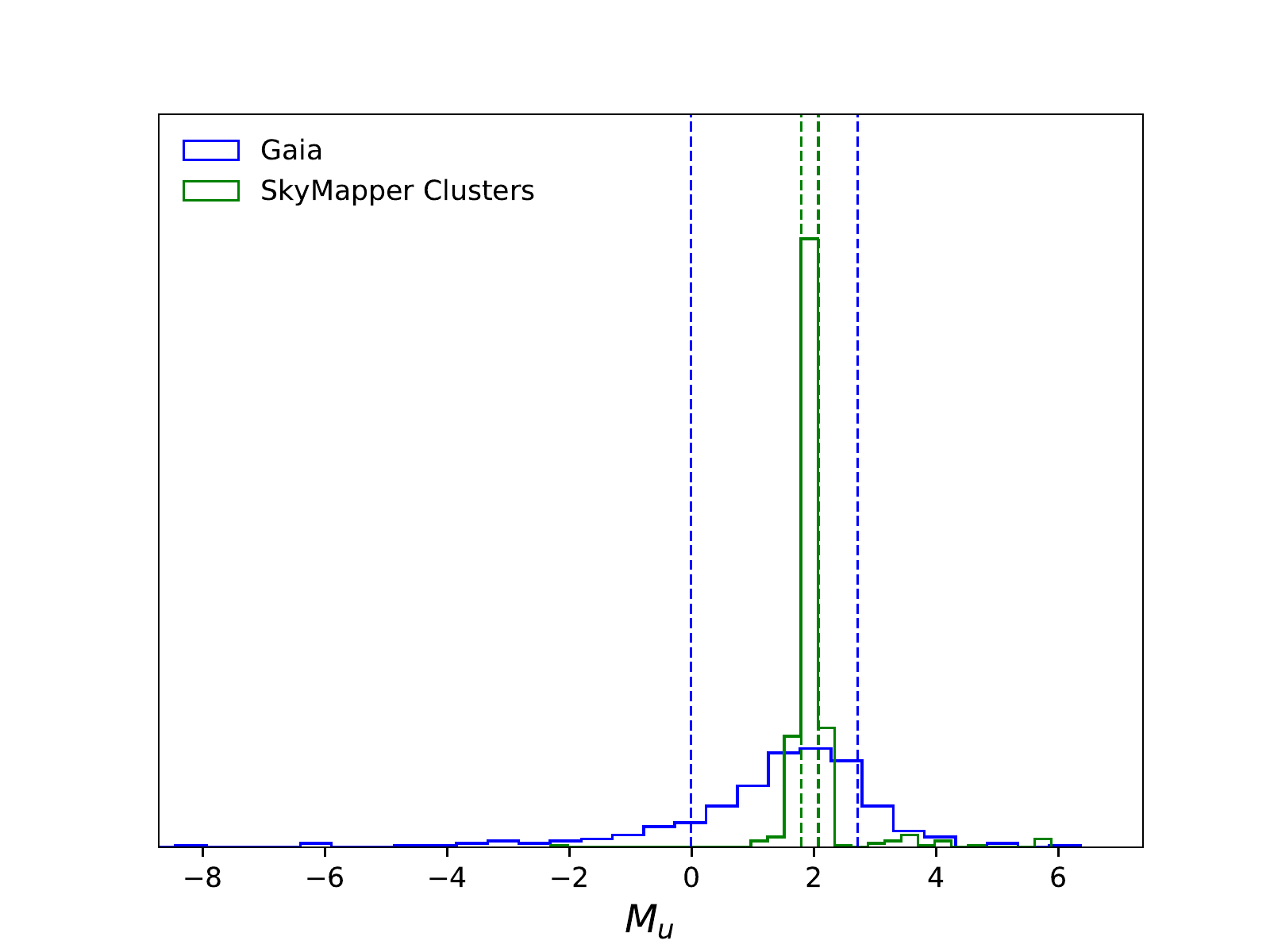}
  \caption{The distribution of stellar absolute magnitudes 
derived from Gaia parallaxes (blue) and from the cluster calibration presented in this paper (green).  
Both distributions peak at  $\sim 2 \magt$, while the results from Gaia possess a larger spread ($\sigma \sim 1 \magt$) due to parallax uncertainties. The $16^{th}$ to $84^{th}$ distribution intervals are indicated with dashed lines. }
  \label{Fig:Gaiaresult}
\end{figure}
\begin{figure}
    \includegraphics[width=\columnwidth]{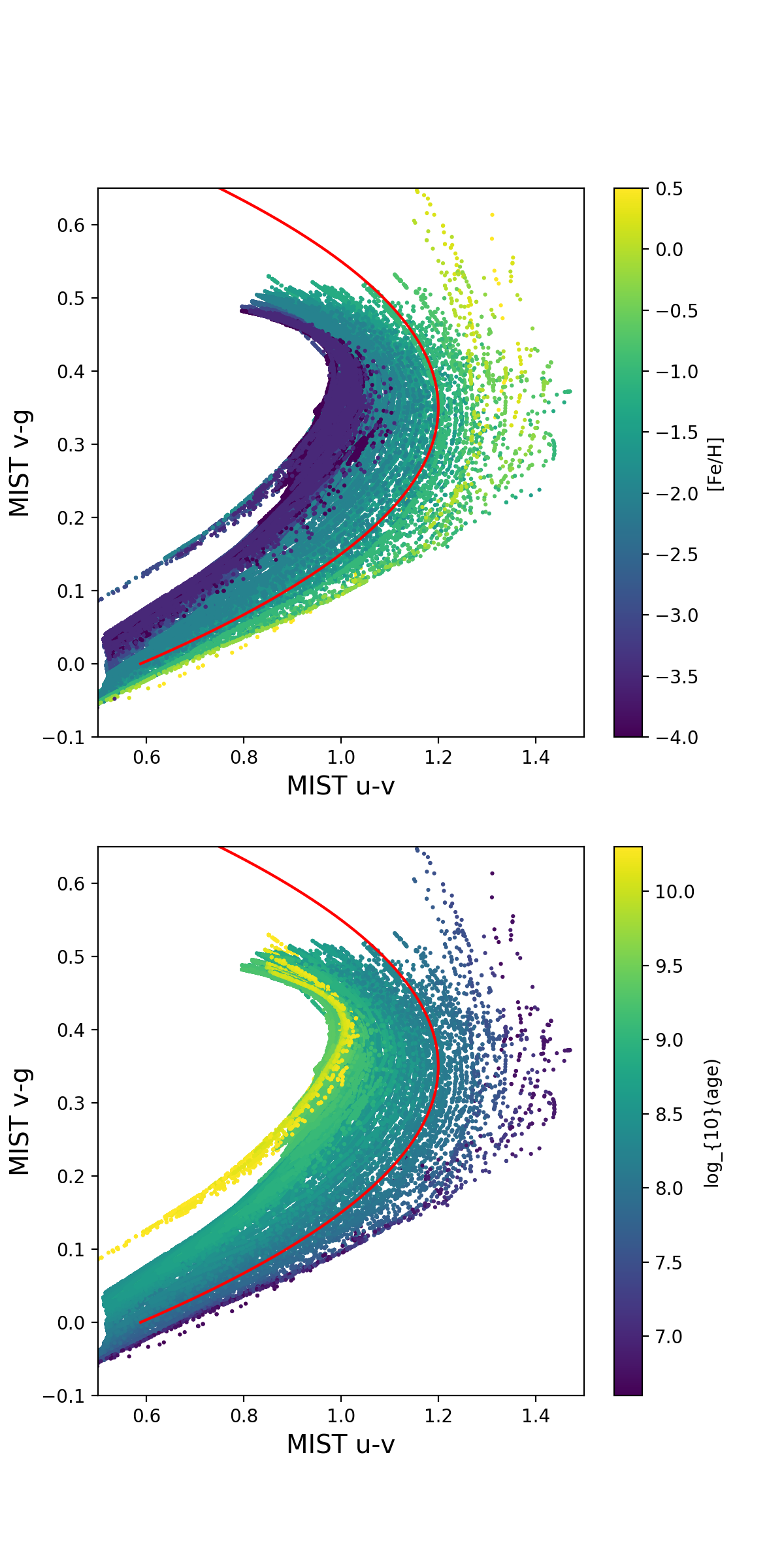}
        \caption{The colour-colour distributions of 
        BHB stars drawn from MIST isochrones with $-4 < \FeH < 0.5$, colour-coded in terms of metallicity (upper panel) and age (lower panel);
these clearly reveal colour-metallicity/colour-age correlations. Metal-poor/old stars are bluer $(u-v)_0$ colour, while metal-rich/young  stars are redder. To represent the observed correlation, we define the quantity, $(u-v)_0 + 5*((v-g)_0 - 0.35)^2$ represented as a red line in both panels.}
        \label{Fig:MISTColourAge}
\end{figure}
\begin{figure}
    \includegraphics[width=\columnwidth]{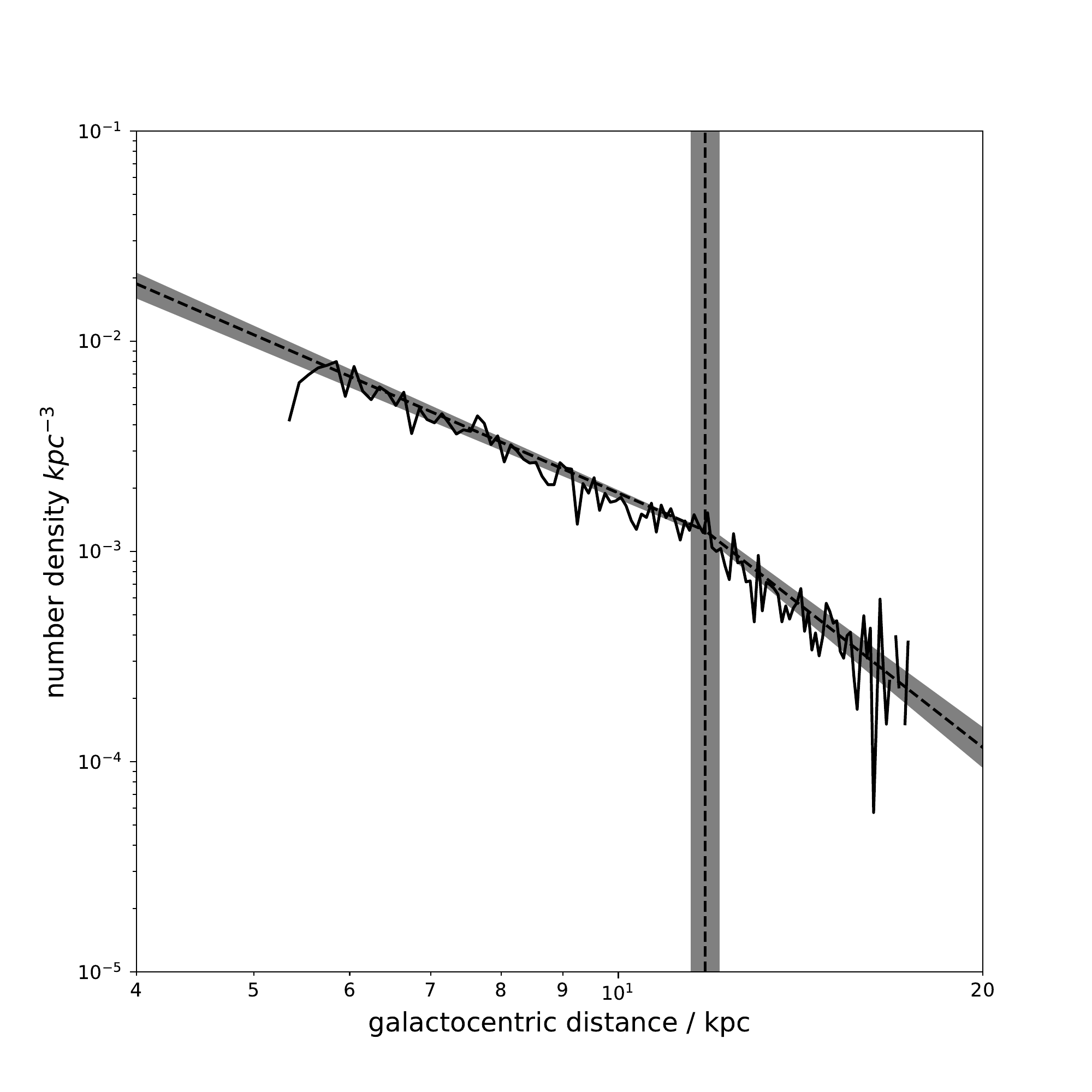}
        \caption{The number density of BHB stars as a function of galactocentric distance. Dashed lines represent the best fitting parameters, whereas the grey regions represent the uncertainty of this fit. We identify a break at $r_s = 11.8\pm0.3\ \kpc$, and within the break radius, the power law index is $\alpha_{in} = -2.5\pm0.1$. Outside of this radius, the power law index is $\alpha_{out} = -4.5\pm0.3$}\label{fig:Numberdensity}
\end{figure}
\begin{figure*}
    \includegraphics[width=\linewidth]{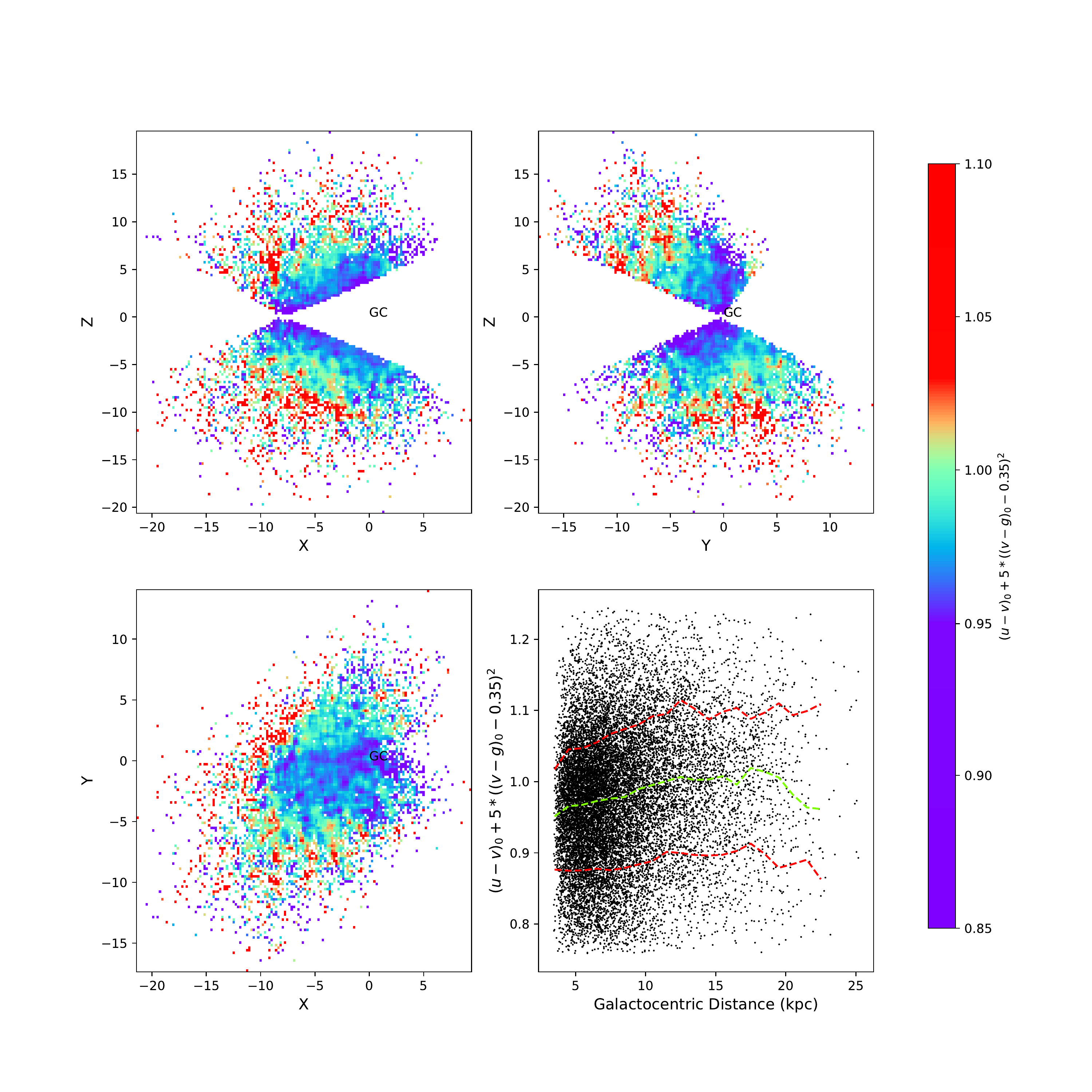}
        \caption{The stellar halo distribution of the quantity  $(u-v)_0 + 5*((v-g)_0 - 0.35)^2$, taken as indicating age/metallicity variations, with substructures being clearly visible.
Lower (blue-purple) values of this quantity, which represent metal-poor or old populations, are concentrated towards the Galactic Centre with some clusters of younger or metal-rich (coloured red) located at the outskirts. 
In the lower right panel, we present this age/metallicity population indicator vs galactocentric distance, where green dash line represents the averaged value in corresponding distance range, and red dash lines represent the $16^{th}$ and $84^{th}$ percentiles. An upward trajectory is clear from $\sim 0.95$ at $\sim 3 \kpc$ to $\sim 1$ at $10 \kpc$.}\label{fig:Final_res}
\end{figure*}

Equation~\ref{eq:linear_part}, as well as the sky position, locate each BHB star in 3-dimensional space. We present the heliocentric radial number distribution of the BHB stars from our sample in \autoref{Fig:Numberdistribution}, where they extend to $\sim 11 \kpc$. 

\subsection{Scaling from Gaia DR1 Parallax}
Parallax is the most straightforward way to determine distance, which has been part of {\it Gaia} project \citep{Gaia2016a, Gaia2016b}. 
Given the overlap between {\it Gaia} and the \smp\ footprints, we searched for BHB stars with parallax measured by {\it Gaia} DR1, and further determine the absolute magnitude for those BHB stars. This acts as an independent measurement, complementing that from the clusters calibration.

We found 270 BHB stars with parallaxes listed in {\it Gaia} DR1, combined \smp\ photometry, whose absolute magnitudes were calculated with:
\begin{gather}
  M = m + 5\log_{10}(0.01*\varpi) \notag \\
  \sigma_{M} = 5 \frac{\varpi}{ln10}\times \sigma_{\varpi}
  \label{absmag}
\end{gather}
The distribution of the absolute magnitude of BHB stars calibrated with {\it Gaia} parallax and clusters are presented in \autoref{Fig:Gaiaresult}. We find a single peak at $\sim 2 \magt$ for both distributions while the result from {\it Gaia} disperses relatively larger due to parallax uncertainties. From \autoref{absmag}, we find the absolute magnitude uncertainty from parallax is $\sigma_{m_0} = 1.46 \magt$. No prominent second peak means that there is no distinct different population of stars in our select BHB sample set.

\section{Colour-colour relationship for BHB stars}
\label{sec:age_cal}
While their luminosities will be roughly constant, the temperature, and hence the colour of old BHB stars will be influenced by their metallicity and envelope mass: metal-rich BHB stars will be cooler since it increases the opacity of the stellar envelope; different envelope masses, resulted from processes such as mass loss in the RGB phase, will further vary the temperature of a population of BHB stars.

To investigate how the physical properties of BHB stars influence their photometric properties, previous studies employ stellar-evolution code to derive the correlation between BHB colour and age \citep[e.g.][ with the MESA isochrone \citep{Paxton2011, 2013ApJS..208....4P, 2015ApJS..220...15P}]{Santucci2015}. 
For the work presented here, we generate a series of MIST \citep[MESA Isochrone and Stellar Track,][]{Dotter2016, Choi2016} isochrone from $\FeH = -4$ to $\FeH = 0.5$ with Kroupa initial mass function \citep[IMF;][]{2001MNRAS.322..231K}. We select those stars that are:

\begin{itemize}
\item[-]On the core helium burning branch
\item[-]Core mass larger than 10\% of stellar mass 
\item[-]Effective temperature of $7000K < T <13000K$
\end{itemize}

These selections exclude massive core helium burning stars and red horizontal branch stars. In \autoref{Fig:MISTColourAge} we present the selected MIST stars in terms of $(u-v)_0$ vs $(v-g)_0$ colours, 
colour-coded with metallicity (upper panel) and age (lower panel) respectively. 
These demonstrate that $(u-v)_0$ colour is sensitive to the properties of the BHB stars, with metal-poor/older stars being bluer in $(u-v)_0$ colour, but clearly age and metallicty are degenerate.

Given that there is not a simple relationship for BHB stars in this colour-colour space, we consider a simple re-parametrization of the properties seen in 
\autoref{Fig:MISTColourAge}. For this, we define the quantity $(u-v)_0 + 5*((v-g)_0 - 0.35)^2$, to indicate the different properties, with the red line in each panel denoting a fiducial case where $(u-v)_0 = -5*((v-g)_0 - 0.35)^2 + 1.2$.
A larger value of this quantity indicates more metal-rich/younger BHBs, while conversely a smaller value indicates a metal-poor/older population. This relationship will be used in the discussion of the results of this paper, to show metallicity/age variability through the stellar halo as seen in the southern sky.

\section{Results}\label{sec:result}

\subsection{Number Density Distribution}\label{NumDen}
The stellar number density distribution of the Galactic halo is thought to trace the accretion history of the Galaxy, and there has been a number of key recently studies using different indicators to depict the profile.
For inner Galactic halo, \citet{Xue2015} used RGB stars to find a power-law index $\alpha_{in} = 2.1 \pm 0.3$ and flattening $q_{in} = 0.70 \pm 0.02$, with a break radius $r_s = 18 \pm 1 \kpc$; \citet{PilaD2015} found a steeper $\alpha_{in} = 2.5 \pm 0.04$, $q_{in} = 0.79 \pm 0.02$ ($r_s = 19.5 \pm 0.4 \kpc$) with MSTO stars; a even steeper stellar halo of $\alpha_{in} = 2.8 \pm 0.4$, $q_{in} = 0.7$ (fixed) ($r_s = 28.5 \pm 5.6 \kpc$) were found by \citet{Faccioli2014} with RR Lyrae Stars; BHB stars were used as a tracer by \citet{2011MNRAS.416.2903D}, which found an intermediate $\alpha_{in} = 2.3 \pm 0.1$, $q_{in} = 0.59 \pm 0.03$ ($r_s = 27.1 \pm 1 \kpc$).

It is insightful to examine the BHB halo number density with our sample derived from \autoref{Fig:Numberdistribution}. To do this, we first select stars with Galactic altitude $|Z| > 4 \ \kpc$ to avoid contamination from the disk and $u < 17.1 \magt$ to avoid incompleteness issues. This finally selects 5272 BHB stars, with which we calculate the stellar density by
\begin{gather}
	\rho(r) = N(r)/\epsilon(r) \notag \\
	r^2 = X^2 + Y^2 + \left(\frac{Z}{q}\right)^2
    \label{func:number_density}
\end{gather}
where $N(r)$ is the number of stars at $r$, $\epsilon(r)$ is the surface area at radius $r$ covered by our sample, and $X$, $Y$ and $Z$ are galactocentric coordinates of the stars, and $q$ represents a flattening of the distribution.

With \autoref{func:number_density}, we find a break in the number density profile around $r = 11\ \kpc$; to depict the profile, we run an MCMC routine \citep{2013PASP..125..306F} with a double power law and a fixed $q = 0.76$ (consistent with star counts at high galactic latitude \citep[see][ and reference therein]{Sharma2011}).
\autoref{fig:Numberdensity} demonstrates the profile and the MCMC results: the best fit break radius is $r_s = 11.8\pm0.3\ \kpc$; within the break radius, the power law index is $\alpha_{in} = -2.5\pm0.1$, while the index beyond the break radius is $\alpha_{out} = -4.5\pm0.3$.

\subsection{Colour Distribution}\label{sec:ColourDistribution}
In the following, we present the colour (the redefined quantity in \autoref{sec:age_cal}) distribution of BHB stars in the Galactic halo as revealed by \smp, in particular focusing upon the large-scale variations and inhomogeneities in the BHB population.

To obtain an overview of the colour distribution, we create a binned map with the bin size of $1 \kpc^2$, and apply a Gaussian smoothing with kernel $\sigma = 2\ $ pixel length. In \autoref{fig:Final_res}
we present the distributions in the X-Z, X-Y and Y-Z Galactic planes: most stars in the centre are relatively blue, which are either old or metal-poor; superimposed upon this background, colour fluctuations are clear---those substructures are relatively young or metal-rich.

With a rough estimation from \autoref{Fig:MISTColourAge}, a change of $0.1\ \magt$ in colour corresponds to a change of $\sim 3-4\ \Gyr$ in age or $\sim 0.5\ dex$ in $\FeH$. That implies that in the concentration of old/metal-poor stars, there still is an age fluctuation $\sim1\ \Gyr$ or metallicity fluctuation of $\sim0.1\ dex$, with a scale of $\sim1\ \kpc$. 

The lower right panel of \autoref{fig:Final_res} presents the distance-colour diagram, where we saw a colour shift of $0.05 \magt$ in the distance range $0-15\ \kpc$, indicating a systematic metallicity ($\sim 0.2 dex$) or age ($\sim 1-2\ \Gyr$) change outwards, and in agreement with the results from \citet{Preston1991}.

\section{Discussion and Conclusions} \label{sec:conclusion}
\smp\ is providing us with a new view of the southern sky, with which, in this paper, we present a new colour map of the Galactic halo distribution of BHB stars.
The BHB stars in the \smp\ colour-colour diagram are separated from other stars so that it is easier to isolate the target stars with the ANN, which is more straightforward than that applicable to SDSS since \smp\ has a bluer and narrower filter set. With this, future data releases of \smp\ offer great promise for providing a precise number density profile and detailed maps of the colour-age/metallicity distribution of the halo. Using the current version of \smp\, we have reached following conclusions with BHB stars:

We calculate the absolute magnitudes of BHB stars based upon several well measured globular clusters. These span a narrow range $\sim\ M_u = 2\ \magt$, varying slightly with colour; and the dispersion of the magnitude of BHB stars is $0.124\ \magt$. Incorporating our adopted systematic distance uncertainty of $5\%$, the overall distance uncertainty increases to $\sim8\%$. This change does not significantly influence the conclusions of this paper, with very little influence on the colour distributions and other presented properties. We note for completeness that
the largest resulting impact is an increase of the uncertainty of the outer power law index by $\sim0.1$.

As a comparison, we also derive the absolute magnitude of a subset of BHB stars with a measured parallax from {\it Gaia} DR1. Though they have a different dispersion, we find these two methods give the same result within the uncertainties. Since the averaged absolute magnitude of BS stars is $\sim2\ \magt$ fainter than BHB stars \citep{2011MNRAS.416.2903D}, we expect the absolute magnitude of BS stars from {\it Gaia} should peaks at $M_u = 0$. As no such peak is apparent, it is clear that our selection has produced a relatively clean sample of BHB stars.

The BHB properties, in particular, mass, metallicity and age will influence its colour, and 
to consider the colour-colour properties of BHB stars, we generated synthetic stellar samples using the MIST isochrones.
From these, it is clear that the colour influences due to age and/or metallicity cannot be uniquely distinguished from photometric data only. Our conclusions, therefore, reflect potential variation in either, or both, of these quantities.

We estimate the halo stellar number density based on BHB stars; found a break at galactocentric distance $r_s = 11.7\pm0.3 \kpc$. Within the break radius, we found a power law index of $\alpha_{in} = -2.5\pm0.1$, while the index beyond the break radius is $\alpha_{out} = -4.5\pm0.3$. 
The inner index agrees with previous results, but the break radius is much smaller than previous claims of $r_s = 25 \pm 10 \kpc$ \citep[e.g.][]{Bland-Hawthorn2016, Xue2015, 2011MNRAS.416.2903D, PilaD2015}. \citet{2018arXiv180107834W} suggests the median point source completeness limits is $u < 17.75$. While we set the magnitude cut at $u < 17.1$ to avoid the impact of incompleteness, the full survey photometric limits and zero-points of the \smp\ Survey will be investigated in later data releases.

Finally, we present a 3-dimensional colour map of BHB stars in the southern sky, revealing substantial substructures that indicate significant age $(\sim1\ Gyr)$ or metallicity $(\sim0.1\ dex)$ variations which are important as they are the potential signatures of the accretion history of the Galaxy \citep{Bullock2005}; accompanying those substructures, we find a systematic colour shift from the centre of the Milky Way outwards, suggesting a large-scale metallicity/age variation through the halo.
Such variations are natural predictions of an accreted stellar halo in hierarchical cosmological formation models \citep{Bullock2005}.
This result resembles some previous works' conclusions:
most recently, \citet{Grady2018} found a similar age gradient with O-Mira stars; 
\citet{Carollo2016} and \citet{Santucci2015} presented the BHB colour distribution in the northern sky based on SDSS, interpreting these substructures as age fluctuations; \citet{Ibata2009} found significant small-scale variations of colour and metallicity in NGC 891, an edge-on galaxy that is an analogue of the Milky Way. \citet{Font2006} suggests that in their simulation, most metal-poor stars in the Galactic halo are buried within the central $\sim 5\ \kpc$ of the Galaxy, indicating that BHB colour substructure could also be resulted from metallicity variation.

It also clear that the \smp\ DR1 is not yet deep enough for a thorough exploration of the Galactic halo. Assuming that BHB stars possess an absolute magnitude of $M_u = 2\ \magt$, based on Sec.\ref{sec:Distance_Cal}, a BHB star with an apparent magnitude $u_0 = 17.75\ \magt$ will be at a distance of $14.12\ \kpc$. However, \smp\ clearly has the advantage of a narrower filter set to select a better BHB sample and future data releases hold great promise for expanding our understanding of the Galactic halo.

\section*{Acknowledgements}
ZW gratefully acknowledges financial support through a the Dean's International Postgraduate Research Scholarship from the Physics School of the University of Sydney.
ADM is grateful for support from an ARC Future Fellowship (FT160100206).
GFL thanks the University of Surrey for hosting him as an IAS fellow for the final stages of the 
preparation of this paper.

The national facility capability for \smp\ has been funded through ARC LIEF grant LE130100104 from the Australian Research Council, awarded to the University of Sydney, the Australian National University, Swinburne University of Technology, the University of Queensland, the University of Western Australia, the University of Melbourne, Curtin University of Technology, Monash University and the Australian Astronomical Observatory. \smp\ is owned and operated by The Australian National University's Research School of Astronomy and Astrophysics. The survey data were processed and provided by the \smp\ Team at ANU. The \smp\ node of the All-Sky Virtual Observatory (ASVO) is hosted at the National Computational Infrastructure (NCI). Development and support the \smp\ node of the ASVO has been funded in part by Astronomy Australia Limited (AAL) and the Australian Government through the Commonwealth's Education Investment Fund (EIF) and National Collaborative Research Infrastructure Strategy (NCRIS), particularly the National eResearch Collaboration Tools and Resources (NeCTAR) and the Australian National Data Service Projects (ANDS).

This work has made use of data from the European Space Agency (ESA)
mission {\it Gaia} (\url{https://www.cosmos.esa.int/Gaia}), processed by
the {\it Gaia} Data Processing and Analysis Consortium (DPAC,
\url{https://www.cosmos.esa.int/web/Gaia/dpac/consortium}). Funding
for the DPAC has been provided by national institutions, in particular
the institutions participating in the {\it Gaia} Multilateral Agreement.

Software credit:
{\sc sklearn} \citep{scikit-learn}, {\sc ipython} \citep{ipython}, {\sc matplotlib} \citep{matplotlib}, {\sc numpy} \citep{numpy}, {\sc pandas} \citep{pandas}, {\sc emcee} \citep{2013PASP..125..306F}




\bibliographystyle{mnras}
\bibliography{zw} 

\bsp    
\label{lastpage}
\end{document}